\documentclass[floatfix,reprint,aip,jcp,amsmath,citeautoscript]{revtex4-1}

\usepackage{graphicx}
\usepackage{braket}
\usepackage{txfonts}
\usepackage{verbatim}
\usepackage{bm}
\usepackage{color}
\definecolor{myblue}{rgb}{0,0,1}
\usepackage[breaklinks=true,colorlinks=true,linkcolor=myblue,urlcolor=myblue,citecolor=myblue]{hyperref}

\newcommand{\vc}[1]{{\bm{#1}}}

\newcommand{\varN}{{30}}
\newcommand{\varJ}{{1.0}}
\newcommand{\vargsq}{{5.0}}
\newcommand{\varomega}{{0.2}}
\newcommand{\varT}{{1.0}}

\newcommand{\varsamp}{{9000}}

\newcommand{\sign}{\text{sgn}}
\begin{document}

\title{A Reciprocal-Space Formulation of Surface Hopping}

\author{Alex Krotz}
\author{Roel Tempelaar}
\email{roel.tempelaar@northwestern.edu}

\affiliation{Department of Chemistry, Northwestern University, 2145 Sheridan Road, Evanston, Illinois 60208, USA}

\begin{abstract}
Surface hopping has seen great success in describing molecular phenomena where electronic excitations tend to be localized, but its application to materials with band-like electronic properties has remained limited. Here, we derive a formulation of fewest-switches surface hopping where both the quantum and classical equations of motion are solved entirely in terms of reciprocal-space coordinates. The resulting method is directly compatible with band structure calculations, and allows for the efficient description of band-like phenomena by means of a truncation of the Brillouin zone. Using the Holstein and Peierls models as examples, we demonstrate the formal equivalence between real-space and reciprocal-space surface hopping, and assess their accuracy against mean-field mixed quantum--classical dynamics and numerically-exact results.
\end{abstract}
 
\maketitle

\section{Introduction}\label{sec:intro}

The recent years have seen a surge in interest in crystalline materials where band-like electronic conductance coexists with strong electron--phonon interactions, examples of which include hybrid metal-halide perovskites \cite{kojima2009organometal, zhu2015charge, wright2016electron} and monolayer transition-metal dichalcogenides \cite{mak2010atomically, splendiani2010emerging, shree2018observation, trovatello2020strongly, li2021exciton}. Few of the theoretical methods currently available are able to efficiently describe lattice-based electron--phonon couplings beyond the perturbative regime \cite{huang2017polaron, lengers2020theory, brem2020phonon}, hampering our ability to unravel the nonequilibrium behavior of such materials.

We recently proposed a formulation of mixed quantum--classical dynamics tailored to band-like models \cite{krotz2021reciprocal}. Mixed quantum--classical dynamics has seen great success in non-perturbatively describing strong interactions between local electronic excitations and nuclear vibrations in molecular systems \cite{subotnik2016understanding, crespo-otero2018recent, nelson2020nonadiabatic}, and is conventionally performed by solving the involved quantum and classical equations of motion in real space. Our formulation instead expresses a periodic lattice of classical harmonic modes with energy $\omega$, positions $q_n$, and momenta $p_n$ in terms of the normal coordinates
\begin{equation}
    z_n\equiv\sqrt{\frac{\omega}{2}}\Big(q_n+i\frac{p_n}{\omega}\Big),
\end{equation}
after which a discrete Fourier transform yields a reciprocal-space representation for both quantum and classical degrees of freedom, including the associated equations of motion. Here, $n=1,\ldots,N$ runs over the lattice sites. Applying a mean-field (MF) description of the quantum--classical coupling terms, the resulting dynamics was previously shown to be identical to that provided by the conventional real-space formulation \cite{krotz2021reciprocal}. Importantly, the method uniquely allows for truncating the quantum and classical bases by harnessing a ``localization'' of excitations in the Brillouin zone while retaining accurate results, which may yield significant computational cost savings \cite{krotz2021reciprocal}. Combined with a direct compatibility with band structure calculations, this renders such a reciprocal-space formulation an ideal avenue for extending the success of mixed quantum--classical dynamics to crystalline materials.

Here, we introduce a reciprocal-space formulation of surface hopping. We specifically focus on fewest-switches surface hopping (FSSH) \cite{tully1990molecular}, which has emerged as one of the most popular variants of mixed quantum-classical dynamics, and which is known to improve over MF (or Ehrenfest) dynamics in its ability to accurately describe detailed balance \cite{tully1998mixed, parandekar2005mixed}. We present its formulation wherein the equations-of-motion are solved fully within reciprocal space (while associating the surfaces with instantaneous quantum eigenstates, as per convention). Similarly to our previous findings for MF dynamics \cite{krotz2021reciprocal}, reciprocal-space FSSH is shown to yield identical results compared to the real-space equivalent. However, compared to MF dynamics, its improved thermalization behavior affects the degree at which Brillouin zone ``localization'' can be harnessed by basis truncations. We apply the method to the Holstein and Peierls models involving a single electronic carrier coupled to a single dispersionless phonon mode, discuss the short-time behavior as well as thermalization properties, and draw a comparison with MF dynamics and exact results.

\section{Theory}\label{sec:theory}

\subsection{Surface hopping in real space}\label{sec:real}

Before turning to reciprocal-space FSSH, we will first summarize the key aspects of the method formulated in the real-space bases. This is not intended as a comprehensive introduction to surface hopping, for which excellent readings can be found in the literature \cite{hammesschiffer1994proton, subotnik2016understanding, wang2016recent, crespo-otero2018recent, nelson2020nonadiabatic}. We start with the total Hamiltonian for the coupled electron--phonon system. Within the classical approximation for the phonon degrees of freedom, this Hamiltonian is partitioned as
\begin{equation}
    \hat{H} = \hat{H}_\text{el} +  \hat{H}_\text{el--ph}(\vc{q}) + H_\text{ph}(\vc{q},\vc{p}).
\end{equation}
Here, the operators $\hat{H}_\text{el}$ and $\hat{H}_\text{el--ph}(\vc{q})$ account for the purely-electronic and electron--phonon coupling contributions, where the latter depends parametrically on the classical position vector, $\vc{q}=(q_1,q_2,\ldots,q_N$). $H_\text{ph}(\vc{q},\vc{p})$ is the classical Hamiltonian accounting for the phonon energy, depending on both position and momentum vectors. Instantaneous quantum eigenstates are found by solving the Schr\"odinger equation
\begin{equation}
    \hat{H}\ket{\alpha} = \big(\hat{H}_\text{el} +  \hat{H}_\text{el--ph}(\vc{q})\big)\ket{\alpha} = \epsilon_\alpha\ket{\alpha},
\end{equation}
with $\epsilon_\alpha$ as the associated eigenenergy. Note that this energy inherits the parametric dependence on $\vc{q}$, as does the eigenvector $\ket{\alpha}$.

Mixed quantum--classical dynamics amounts to self-consistently solving the coupled quantum and classical equations of motion. The latter are given by Hamilton's equations
\begin{equation}
    \dot{q}_n = \frac{\partial H_\text{ph}}{\partial p_n}, \quad \dot{p}_n = -\frac{\partial H_\text{ph}}{\partial q_n} -\frac{\partial \langle H_\text{el--ph}\rangle}{\partial q_n}.
\end{equation}
Here, $\langle H_\text{el--ph}\rangle$ is the expectation value of the electron--phonon coupling operator. This expectation value is determined most straightforwardly through the MF expression
    $\langle H_\text{el--ph}\rangle = \braket{\Psi|\hat{H}_\text{el--ph}|\Psi},$
with $\ket{\Psi}$ as the electronic wavefunction propagated through the time-dependent Schr\"odinger equation ($\hbar=1$ is taken here and throughout)
\begin{equation}
    i\ket{\dot{\Psi}} = \big(\hat{H}_\text{el} +  \hat{H}_\text{el--ph} \big)\ket{\Psi}.
    \label{eq_time_dep_SE}
\end{equation}
The electronic wavefunction can be expanded in the eigenbasis as
\begin{equation}
    \ket{\Psi}=\sum_\alpha A_\alpha \ket{\alpha}.
\end{equation}

Surface hopping, rather than an MF average, associates the electron--phonon coupling expectation value with a single quantum eigenstate, labeled $a$, $\langle H_\text{el--ph}\rangle = \braket{a|\hat{H}_\text{el--ph}|a}$, considered to be an ``active'' surface. This active surface is allowed to stochastically switch among eigenstates. Within FSSH, the probability for a switch to occur is governed by the expression
\begin{equation}
    P_{a:\alpha\rightarrow\beta}=-2\Re\left(\vc{p}\cdot \vc{d}_{\alpha\beta} \frac{A_\beta}{A_\alpha} \right)\Delta t,
\end{equation}
where $\alpha$ and $\beta$ label the initial and terminal eigenstates, and where $\vc{d}_{\alpha\beta}$ is their associated nonadiabatic coupling vector given by
\begin{equation}
    \vc{d}_{\alpha\beta}=\braket{\alpha|\nabla_\vc{q}|\beta}.
\end{equation}
Furthermore, $A_{\alpha(\beta)}$ is the coefficient obtained by propagating the electronic wavefunction through Eq.~\ref{eq_time_dep_SE}, and $\Delta t$ is the time increment for which the switching probability is evaluated.

In addition to the stochastic procedure of switching among quantum eigenstates, the constraint is added that the total (quantum plus classical) energy is conserved. This implies that when a switching from $\alpha$ to $\beta$ occurs, the classical momenta are adjusted such that the classical energy offsets the change in quantum energy, $\epsilon_\beta-\epsilon_\alpha$. Specifically, the momenta are rescaled along the direction of the nonadiabatic coupling vector such that the new momenta $\vc{p}'$ relate to the previous momenta $\vc{p}$ as
\begin{equation}
    \vc{p}' = \vc{p} - \gamma_{\alpha\beta}\vc{d}_{\alpha\beta}.
\end{equation}
Here, $\gamma_{\alpha\beta}$ is obtained by equating the previous total energy to the new total energy \cite{hammesschiffer1994proton}. In case when no sufficient classical kinetic energy is available to accommodate a switch from a lower to higher eigenstate, the switching event is aborted.

Lastly, the electronic density matrix is commonly constructed based on the active surface and the electronic wavefunction coefficients as \cite{tempelaar2013surface, landry2013correct, chen2016on}
\begin{equation}
    \rho_{\alpha\beta}=\delta_{\alpha\beta}\delta_{\alpha a}+(1-\delta_{\alpha\beta})A_\alpha^* A_\beta,
    \label{eq:DM_eigenbasis}
\end{equation}
with $\delta$ as the Kronecker delta function. 

\subsection{Surface hopping in reciprocal space}\label{sec:reciprocal}

\begin{figure}
\includegraphics[scale=.75]{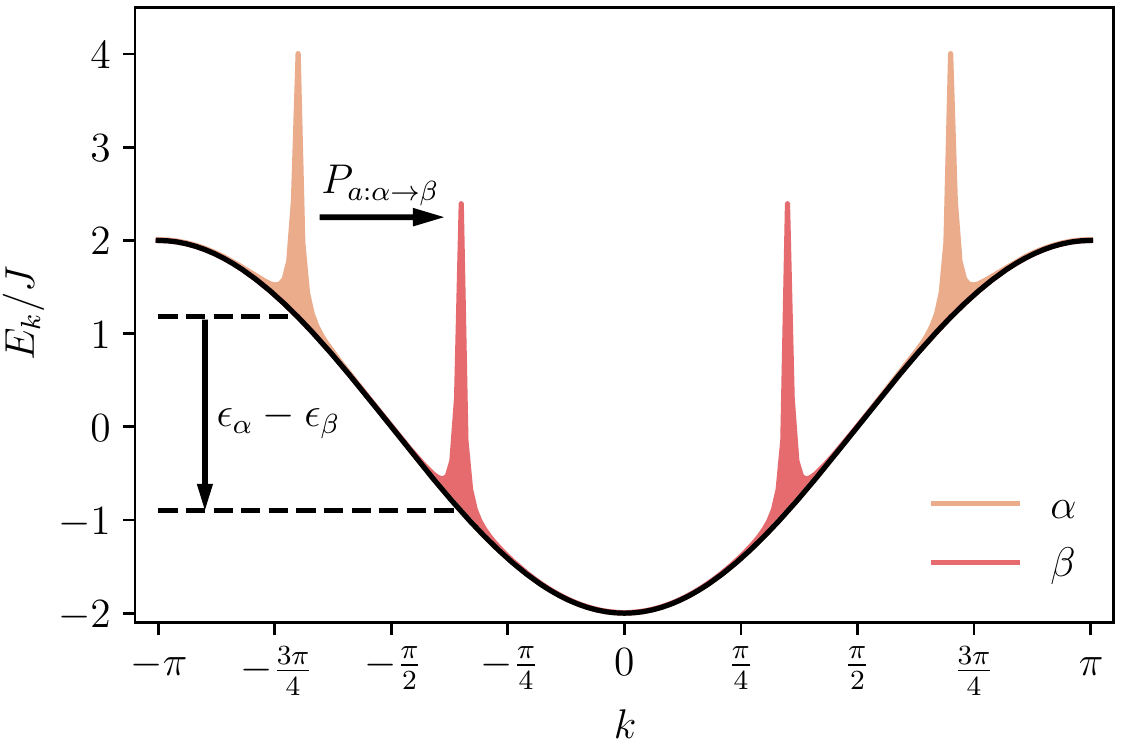}
\caption{Schematic reciprocal-space depiction of a switching event between eigenstates $\alpha$ and $\beta$ associated with a single electronic carrier within a tight-binding model. Black solid curve represents the energy dispersion, $E_k=-2J\cos(k)$. Shown on top of this dispersion are the probability distributions of the eigenstates $\alpha$ (orange) and $\beta$ (red), $\vert \tilde{U}^{\alpha/\beta}_k\vert^2$ (vertically rescaled for clarity of presentation) for a representative snapshot of the phonon coordinates. The associated eigenenergies are indicated by dashes.}
\label{fig:hopping_diagram}
\end{figure}

We now proceed to introduce reciprocal-space FSSH for a periodic lattice. Without loss of generality we will specifically consider a lattice in one dimension where each site involves a single quantum state interacting with a single classical mode. The transformation from real to reciprocal space is then given by
\begin{equation}
    \tilde{f}_k=\frac{1}{\sqrt{N}}\sum_n e^{-ikn} f_n,
\end{equation}
with $f_n=z_n$ for the classical coordinates, $f_n=\phi_n$ for the quantum coordinates, and where $k$ is the wavevector running from $-\pi$ to $\pi$. (Henceforth, a tilde will be used to denote reciprocal-space coordinates.) As mentioned in the Introduction, surfaces will be associated with quantum eigenstates, regardless of the bases in which the classical and quantum coordinates are represented. Within the real-space quantum basis, an eigenstate is expanded as
\begin{equation}
    \ket{\alpha}=\sum_n U_n^\alpha \ket{\phi_n},
\end{equation}
whereas the expansion in terms of reciprocal-space quantum basis states is given by
\begin{equation}
    \ket{\alpha}=\sum_k \tilde{U}_k^\alpha\ket{\tilde{\phi}_k}.
    \label{eq:eigenstate_expand_k}
\end{equation}
Fig.~\ref{fig:hopping_diagram} presents a schematic illustration of a switching event between eigenstates $\alpha$ and $\beta$ associated with a single electronic carrier, represented in reciprocal space. Before turning to the associated switching probability, it is worth noting that nonequilibrium quantum dynamics is generally dictated by the set of nonadiabatic coupling vectors. In real space, these vector exclusively involve position derivatives. However, upon introducing canonical positions and momenta in reciprocal space through \cite{krotz2021reciprocal}
\begin{equation}
    \tilde{z}_k\equiv\sqrt{\frac{\omega}{2}}\Big(\tilde{q}_k+i\frac{\tilde{p}_k}{\omega}\Big),
\end{equation}
both position and momentum derivatives will contribute,
\begin{equation}
    \vc{d}_{\alpha\beta}^{(\vc{\tilde{q}})}\equiv\braket{\alpha|\nabla_\vc{\tilde{q}}|\beta}\neq 0,\quad \vc{d}_{\alpha\beta}^{(\vc{\tilde{p}})}\equiv\braket{\alpha|\nabla_\vc{\tilde{p}}|\beta}\neq 0.
\end{equation}
The reason for this is that the classical positions and momenta become scrambled in the canonical transformation from real to reciprocal space \cite{krotz2021reciprocal},
\begin{eqnarray}
    \tilde{q}_k&
    =\frac{1}{\sqrt{N}}\sum_n\bigg(\tilde{q}_n\cos(kn)-\frac{\tilde{p}_n}{\omega}\sin(kn)\bigg),\nonumber
    \\
    \tilde{p}_k&
    =\frac{\omega}{\sqrt{N}}\sum_n\bigg(\frac{\tilde{p}_n}{\omega}\cos(kn)+\tilde{q}_n\sin(kn)\bigg)
    \label{eq:qk_pk}.
\end{eqnarray}
It is nonetheless straightforward to show that 
\begin{equation}
   \vc{p}\cdot \vc{d}_{\alpha\beta} = \vc{\tilde{p}}\cdot \vc{d}_{\alpha\beta}^{(\vc{\tilde{q}})} -\omega^{2} \vc{\tilde{q}}\cdot \vc{d}_{\alpha\beta}^{(\vc{\tilde{p}})},
\end{equation}
and the reciprocal-space switching probability is simply given by
\begin{equation}
    P_{a:\alpha\rightarrow\beta}=-2\Re\left( \left[ \vc{\tilde{p}}\cdot \vc{d}_{\alpha\beta}^{(\vc{\tilde{q}})} -\omega^{2} \vc{\tilde{q}}\cdot \vc{d}_{\alpha\beta}^{(\vc{\tilde{p}})} \right] \frac{A_\beta}{A_\alpha} \right)\Delta t.
\end{equation}

The scrambling of positions and momenta in the canonical transformation from real to reciprocal space also has ramifications for adjusting the classical coordinates, necessary for offsetting the change in quantum energy when switching between surfaces. In real space only $\vc{p}$ can be rescaled, as a rescaling of $\vc{q}$ would violate the locality inherent to a classical description. The reciprocal-space coordinates $\vc{\tilde{p}}$ and $\vc{\tilde{q}}$ both receive contributions from $\vc{p}$, as a result of which both are to be rescaled along the direction of the relevant nonadiabatic coupling vectors as
\begin{equation}
    \vc{\tilde{p}}' = \vc{\tilde{p}} - \gamma_{\alpha\beta} \vc{d}_{\alpha\beta}^{(\vc{\tilde{q}})}, \quad
    \vc{\tilde{q}}' = \vc{\tilde{q}} + \gamma_{\alpha\beta}\vc{d}_{\alpha\beta}^{(\vc{\tilde{p}})}
    \label{eq:reciprocal_adjust}.
\end{equation}

A third consequence of the canonical transformation from real to reciprocal space is that Hamilton's equations take a form that is different from the real-space equivalent,
\begin{eqnarray}
    \dot{\tilde{q}}_k &=& \frac{\partial H_\text{ph}}{\partial \tilde{p}_k} + \frac{\partial \braket{a| \hat{H}_\text{el--ph}|a}}{\partial \tilde{p}_k},\nonumber \\
    \dot{\tilde{p}}_k &=& -\frac{\partial H_\text{ph}}{\partial \tilde{q}_k} - \frac{\partial \braket{a| \hat{H}_\text{el--ph}|a}}{\partial \tilde{q}_k},
\end{eqnarray}
where Hellmann--Feynman forces due to the active surface act on both momenta and positions.

What sets reciprocal-space mixed quantum--classical dynamics apart from its real-space analog is the possibility of describing band-like phenomena under a truncated Brillouin zone, which may come with considerable computational cost savings \cite{krotz2021reciprocal}. A consequence of such truncation is that the nonadiabatic coupling vectors can no longer be trivially taken to be real valued,
\begin{equation}
    \vc{d}_{\alpha\beta}^{(\vc{\tilde{q}}/\vc{\tilde{p}})}
    = \sum_{k,k'}{\tilde{C}_k^\alpha}{}^*\tilde{C}_{k'}^\beta\braket{\alpha|\nabla_{\vc{\tilde{q}}/\vc{\tilde{p}}}|\beta} \not\in \mathbb{R}.
\end{equation}
When rescaling $\vc{\tilde{p}}$ and $\vc{\tilde{q}}$ under a Brillouin zone truncation, we therefore replace the nonadiabatic coupling vectors in Eq.~\ref{eq:reciprocal_adjust} by their absolute values multiplied by the \textit{sign} of their components along the real axis,
\begin{align}
    \vc{d}_{\alpha\beta}^{(\vc{\tilde{q}}/\vc{\tilde{p}})}
    \leftarrow \left\vert \vc{d}_{\alpha\beta}^{(\vc{\tilde{q}}/\vc{\tilde{p}})}\right\vert \sign\left[\Re\left(\vc{d}_{\alpha\beta}^{(\vc{\tilde{q}}/\vc{\tilde{p}})}\right)\right].
\end{align}
This reduces to the standard rescaling procedure in the limit where the entire Brillouin zone is accounted for, in which case $\vc{d}_{\alpha\beta}^{(\vc{\tilde{q}}/\vc{\tilde{p}})}\in\mathbb{R}$. Importantly, no such replacement is necessary when evaluating the switching probabilities, since one can always utilize the chain rule in order to reformulate this probability as \cite{hammesschiffer1994proton}
\begin{equation}
    P_{a:\alpha\rightarrow\beta}=-2\Re\left(
    \Big\langle \alpha\Big\vert\frac{\partial\beta}{\partial t}\Big\rangle
    \frac{A_\beta}{A_\alpha} \right)\Delta t,
\end{equation}
where $\frac{\partial\beta}{\partial t}$ can be evaluated through discrete differentiation using time step $\Delta t$.

\subsection{Holstein and Peierls models}\label{sec:models}

In the following Section, we will apply the reciprocal-space surface hopping approach to the Holstein and Peierls (or Su--Schrieffer--Heeger \cite{su1979solitons}) models. In both cases, the phonon Hamiltonian takes the form of noninteracting, dispersionless, and harmonic oscillators, expressed in real- and reciprocal-space as
\begin{equation}
    H_\text{ph} = \sum_n \left( \frac{1}{2}p_n^2+\frac{1}{2}\omega^2 q_n^2 \right)
    = \sum_k \left( \frac{1}{2}\tilde{p}_k^2+\frac{1}{2}\omega^2 \tilde{q}_k^2 \right).
\end{equation}
Where the Holstein and Peierls models differ is by the form of the electron--phonon interaction Hamiltonian, $\hat{H}_\text{el--ph}$. As detailed in Ref.~\citenum{krotz2021reciprocal}, this Hamiltonian within the Holstein model takes the real and reciprocal-space forms
\begin{eqnarray}
    \hat{H}_\text{el--ph} &=& g\sqrt{2\omega^{3}}\sum_{n}\hat{c}_{n}^{\dagger}\hat{c}_{n} q_n \\
    &=& \frac{g\sqrt{\omega}}{\sqrt{2N}}\sum_{k,\kappa}\hat{c}_{k+\kappa}^\dagger \hat{c}_k
    \Big(\omega\big(\tilde{q}_{-\kappa}+\tilde{q}_\kappa\big)-i\big(\tilde{p}_{-\kappa}-\tilde{p}_\kappa\big)\Big). \nonumber
\end{eqnarray}
Here, $\hat{c}_n^{(\dagger)}$ and $\hat{c}_k^{(\dagger)}$ are the real- and reciprocal-space electronic annihilation (creation) operators, respectively. The dimensionless parameter $g$ is the electron--phonon coupling constant which is related to the nuclear reorganization energy as $g^{2} \omega$. The corresponding Hellmann--Feynman forces are given by
\begin{eqnarray}
    \frac{\partial \braket{a| \hat{H}_\text{el--ph}|a}}{\partial \tilde{q}_k} &=& g\sqrt{\frac{2\omega^3}{N}}\Re\{C_k^a\},\nonumber\\
    \frac{\partial \braket{a| \hat{H}_\text{el--ph}|a}}{\partial \tilde{p}_k} &=& -g\sqrt{\frac{2\omega}{N}}\Im\{C_k^a\},
\end{eqnarray}
where $C_k^a$ is the autocorrelation function defined as
\begin{align}
    C_k^a \equiv \sum_{k'}\braket{a|\hat{c}_{k'+k}^\dagger \hat{c}_{k'}|a}.
\end{align}

As detailed in Ref.~\citenum{krotz2021reciprocal}, the electron--phonon interaction Hamiltonian within the Peierls model takes the real- and reciprocal-space forms
\begin{eqnarray}
    \hat{H}_\text{el--ph}
    &=& g\sqrt{2\omega^{3}}\sum_{n}\left(\hat{c}^{\dagger}_{n}\hat{c}_{n+1}+\hat{c}^{\dagger}_{n+1}\hat{c}_{n}\right)\left(q_{n}-q_{n+1}\right) \nonumber \\
    &=& \frac{g\sqrt{2\omega}}{\sqrt{N}}\sum_{k,\kappa}\hat{c}_{k+\kappa}^\dagger \hat{c}_k \Big(i\omega\big(\tilde{q}_{-\kappa}+\tilde{q}_\kappa\big)+\big(\tilde{p}_{-\kappa}-\tilde{p}_\kappa\big)\Big)\nonumber\\
    && \qquad \times \big(\sin(k+\kappa)-\sin(k)\big).
\end{eqnarray}
The corresponding Hellmann--Feynman forces are given by
\begin{eqnarray}
    \frac{\partial \braket{a| \hat{H}_\text{el--ph}|a}}{\partial \tilde{q}_k} &=& -g\sqrt{\frac{8\omega^3}{N}}\Im\left\{\bar{C}_k\right\},\nonumber\\
    \frac{\partial \braket{a| \hat{H}_\text{el--ph}|a}}{\partial \tilde{p}_k} &=& -g\sqrt{\frac{8\omega}{N}}\Re\left\{\bar{C}_k\right\},
\end{eqnarray}
with $\bar{C}_k^a$ as the modulated autocorrelation function given by
\begin{align}
    \bar{C}_k^a \equiv \sum_{k'}\braket{a|\hat{c}_{k'+k}^\dagger \hat{c}_{k'}|a}\big(\sin(k'+k)-\sin(k')\big).
\end{align}

Without loss of generality we restrict ourselves to a tight-binding model for the electronic Hamiltonian, which takes the real- and reciprocal-space forms
\begin{eqnarray}
    \hat{H}_\text{el} &=& -J\sum_n \left( \hat{c}_{n+1}^\dagger \hat{c}_n
    + \hat{c}_n^\dagger \hat{c}_{n+1} \right) \nonumber \\
    &=& -2J \sum_k \hat{c}^\dagger_k \hat{c}_k \cos(k),
\end{eqnarray}
and whose dispersion is depicted in Fig.~\ref{fig:hopping_diagram}. We furthermore restrict ourselves to the case of a single electronic carrier, which is initialized in the $k=0$ reciprocal-space basis state,
\begin{equation}
    \ket{\Psi}=\ket{k=0}=\frac{1}{\sqrt{N}}\sum_n\ket{n},
\end{equation}
whereas the classical coordinates are sampled independently from a Boltzmann distribution, as detailed in Ref.~\citenum{krotz2021reciprocal}.

\section{Results}\label{sec:results}

\subsection{Holstein and Peierls dynamics under basis truncations}\label{sec:truncations}

\begin{figure}
\includegraphics[scale=1.0]{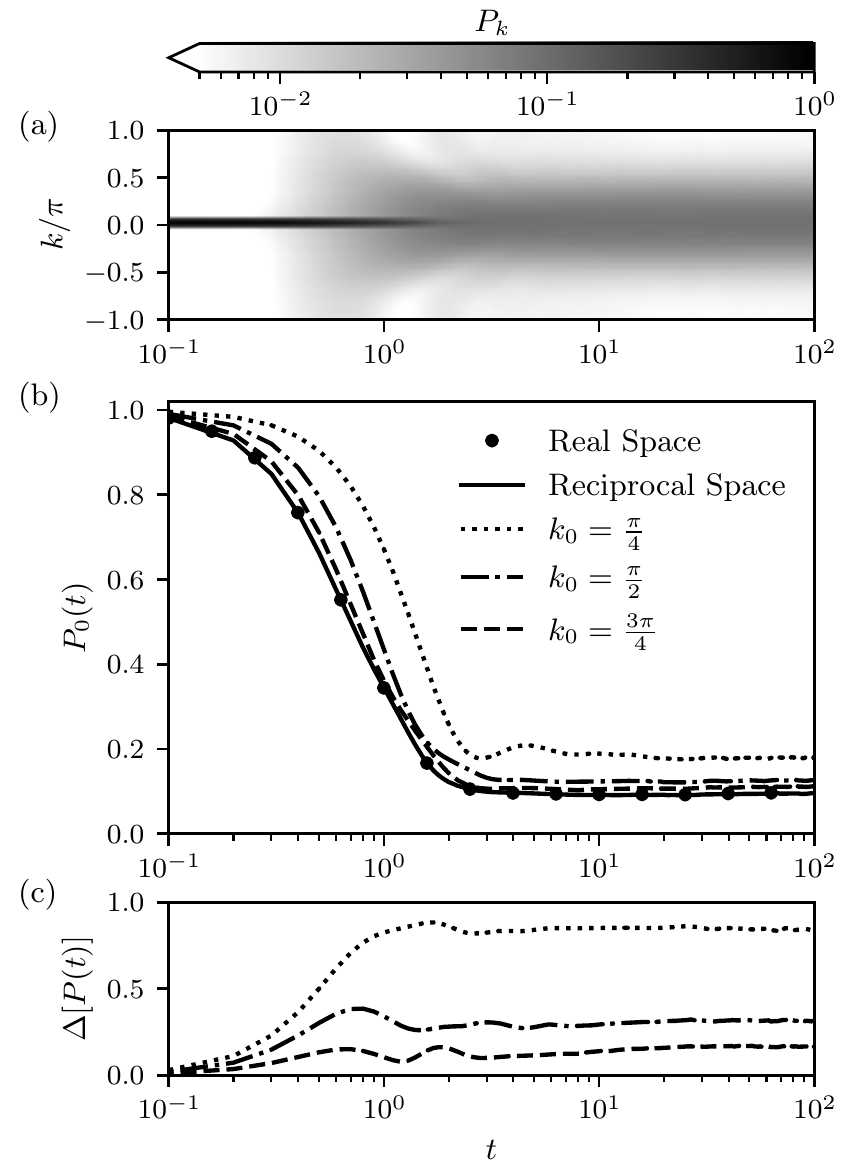}
\caption{Transient electronic populations $P_k(t)$ calculated within FSSH for an $N=30$ site periodic Holstein model with $J=\varJ$, $\omega=\varomega$, $g^2=\vargsq$, and $T=\varT$. Shown in (a) is a heat map of the populations within the first Brillouin zone, obtained through the reciprocal-space formulation of the method. Shown in (b) is $P_0(t)$ obtained through the reciprocal-space (solid curves) and real-space (markers) formulations. For the former, additional results are shown obtained upon truncating the Brillouin zone to $\vert k\vert<k_0$ (dotted, dash-dotted, and dashed curves). Shown in (c) are the cumulative absolute errors relative to the untruncated results. Note the log scale on the horizontal axis.}
\label{fig:results_holstein}
\end{figure}

In this Section we present dynamics generated by reciprocal-space FSSH for the Holstein and Peierls models.
Throughout, a periodic lattice consisting of $N=\varN$ sites is considered with $J=\varJ$, $\omega=\varomega$, $g^{2}=\vargsq$, and $T=\varT$. 
We take the values to be unitless but note that, when taking the thermal energy at room temperature (293~K) as a reference, a unit of energy amounts to 25~meV and a unit of time to 164~fs. We also note that, for the Peierls model, these values are very similar to those commonly used to characterize organic molecular crystals \cite{troisi2006charge}. For each stochastic trajectory the reciprocal-space electronic density matrix is obtained through
\begin{equation}
    \tilde{\rho}_{kl}=\sum_{\alpha,\beta}\tilde{U}_k^\alpha {\rho_{\alpha\beta} \tilde{U}_l^\beta}^*,
\end{equation}
with $\rho_{\alpha\beta}$ being the density matrix in the (instantaneous) eigenbasis given by Eq.~\ref{eq:DM_eigenbasis}, and with $\tilde{U}_k^\alpha$ being the expansion coefficients given by Eq.~\ref{eq:eigenstate_expand_k}. Once transformed, the reciprocal-space electronic density matrix was averaged over $\varsamp$ trajectories. Convergence with respect to the time integration step was assured in all cases.

Shown in Fig.~\ref{fig:results_holstein} are results obtained for the Holstein model with Fig.~\ref{fig:results_holstein} (a) showing time-dependent reciprocal-space electronic populations, $P_k(t)\equiv\tilde{\rho}_{kk}(t)$, obtained through reciprocal-space FSSH. The electronic carrier is seen to rapidly scatter out of the $k=0$ initial state due to interactions with phonons, manifested as a broadening of the reciprocal-space populations, while equilibration sets in at $t\sim5$. In Fig.~\ref{fig:results_holstein} (b) we present transient electronic populations at zero momentum, $P_0(t)$, obtained through FSSH implemented in reciprocal space as well as its conventional implementation in real space. Importantly, and similarly to what was previously observed for MF dynamics \cite{krotz2021reciprocal}, both real- and reciprocal-space implementations yield identical results, confirming their equivalence.

Also shown in Fig.~\ref{fig:results_holstein} (b) are results for which the Brillouin zone was limited to within a truncation radius, $\vert k\vert < k_0$. Deviations in $P_0(t)$ are seen to arise with decreasing value of the radius $k_0$, but remain modest even for truncations as severe as $k_0=\pi/2$ (halving the Brillouin zone). This is further illustrated in Fig.~\ref{fig:results_holstein} (c) which for each truncation radius shows the cumulative absolute error,
$\Delta[P(t)]\equiv\sum_k\vert\Delta P_k(t)\vert$,
where $\Delta P_k(t)$ is the error relative to the untruncated result. (Note that the cumulative absolute error is bounded as $\Delta[P(t)] \in [ 0,2 ]$.) The effectiveness of the applied truncation scheme at early times can be understood based on the initial electronic state being concentrated at $k=0$, whereas its effectiveness in the asymptotic limit is a consequence of the thermalized electronic state being biased to the low-energy region surrounding the same point in the Brillouin zone.

\begin{figure}
\includegraphics[scale=1.0]{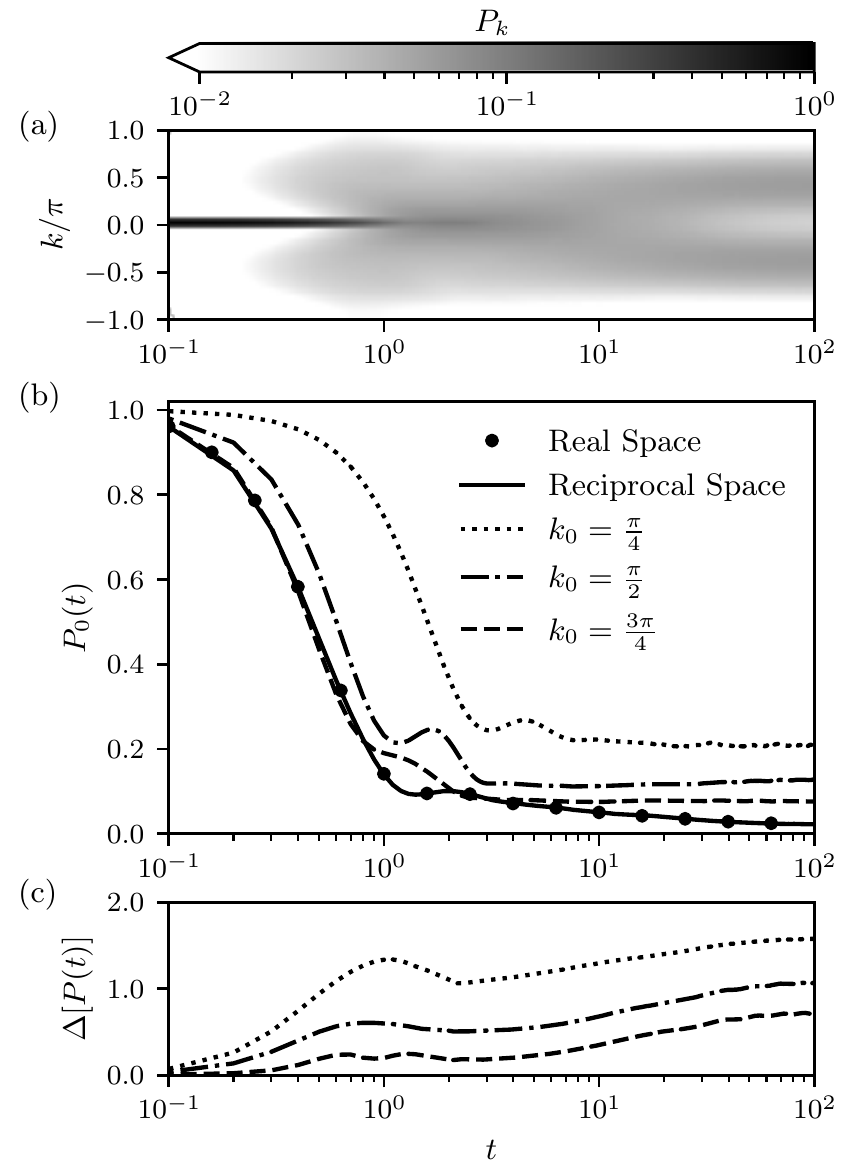}
\caption{Same as Fig.~\ref{fig:results_holstein} but for an $N=30$ site periodic Peierls model with $J=\varJ$, $\omega=\varomega$, $g^2=\vargsq$, and $T=\varT$.}
\label{fig:results_peierls}
\end{figure}

Fig.~\ref{fig:results_peierls} presents results equivalent to those shown in Fig.~\ref{fig:results_holstein}, but for the Peierls model. Importantly, we again find the untruncated reciprocal-space FSSH to generate results identical to the real-space equivalent, consistent with their formal equivalence. Similarly to the Holstein results, interactions with Peierls-coupled phonons yield a rapid broadening of electronic wavepacket. However, equilibration does not set in until $t\sim 50$. Moreover, as seen in Fig.~\ref{fig:results_peierls} (a), at long times the wavepacket is seen to assume a bimodal profile. This behavior is consistent with previous reports \cite{zhao1994on, chen2011on} based on the Munn--Silbey canonical transformation \cite{sibey1980general}, that showed the emergence of new band minima at $k = \pm \pi/2$ which is caused entirely by sizable Peierls coupling. Since the wavepacket is no longer centered at $k=0$, a truncation of the Brillouin zone is markedly less effective than for the Holstein model, yielding comparatively larger errors, as shown in Fig.~\ref{fig:results_peierls} (b) and (c).

\subsection{Surface hopping versus mean-field dynamics}\label{sec:SF_vs_MF}

The parameter values applied in the calculations presented in Sec.~\ref{sec:truncations} are identical to those applied previously in our MF implementation of reciprocal-space mixed quantum--classical dynamics \cite{krotz2021reciprocal}, the results of which are reproduced in the Supplementary Material. Although both implementations can be seen to yield somewhat similar results, there are notable differences. Perhaps most importantly, the MF calculations are characterized by a sustained growth of the cumulative absolute error $\Delta [ P(t) ]$, which is the result of a continual broadening of the electronic wavepacket in reciprocal space. Furthermore, they fail to reproduce the bimodal profile observed for the Peierls model in  Fig.~\ref{fig:results_peierls} (a). As it turns out, these differences are all related to the inability of MF dynamics to account for detailed balance.

In order to assess detailed balance, we determine the effective temperature reached for the quantum subsystem by first constructing a Boltzmann distribution based on the instantaneous quantum eigenstates at the single trajectory level, with the Boltzmann factors given by
\begin{equation}
    s_\alpha(T_\text{eff})=\frac{1}{Z(T_\text{eff})}e^{-\epsilon_\alpha/T_\text{eff}}.
\end{equation}
Here, the effective temperature $T_\text{eff}$ enters as a free parameter, $Z(T_\text{eff})$ is the partition function, and the Boltzmann constant is taken to be unity. The Boltzmann factors are then transformed to the reciprocal-space basis upon which an ensemble average yields a $T_\text{eff}$-dependent distribution $\langle s_k\rangle (T_\text{eff})$. The effective temperature is then obtained by fitting this distribution to the actual populations at a certain instant $t$, $P(t)$.

\begin{figure}
\includegraphics[scale=1.0]{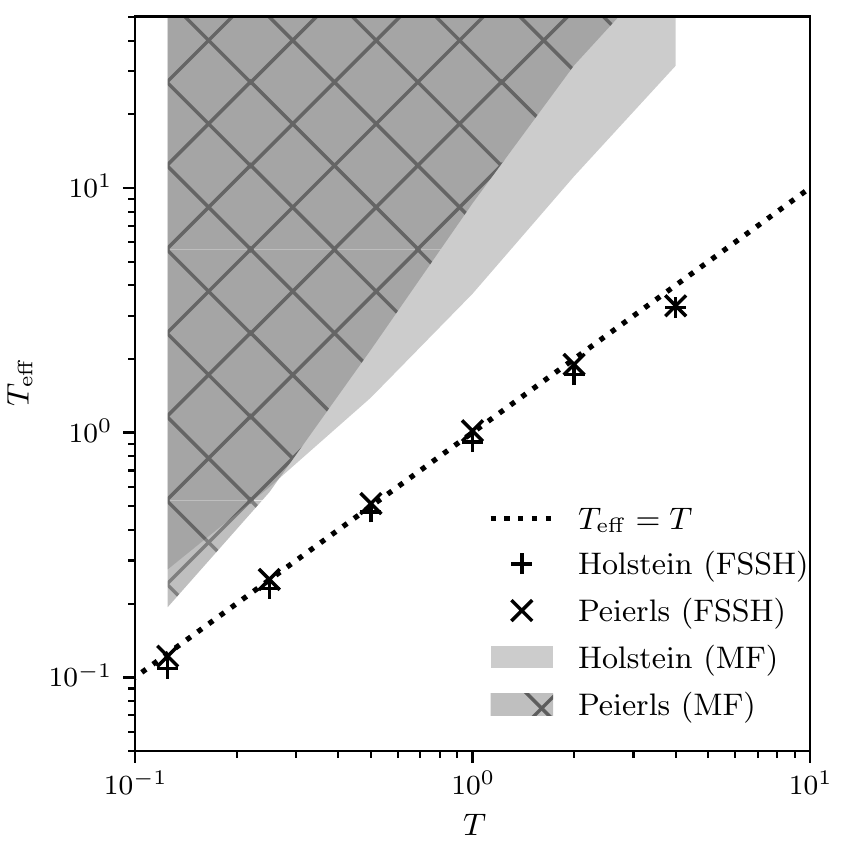}
\caption{Thermalization behaviors of FSSH and MF dynamics for the Holstein and Peierls models for the same parameters as in Figs.~\ref{fig:results_holstein} and \ref{fig:results_peierls}. Shown is the effective temperature of the quantum subsystem $T_\text{eff}$ as a function of the classical temperature $T$. Detailed balance is satisfied when $T_\text{eff}=T$ (dotted line). For MF, only a lower bound to $T_\text{eff}$ could be obtained, and its range of possible values is indicated by the shaded area. In all cases, $T_\text{eff}$ was determined at time $t=100$.}
\label{fig:temp_scan_holstein}
\end{figure}

Shown in Fig.~\ref{fig:temp_scan_holstein} are fitted values of $T_\text{eff}$ against the temperature used to sample the classical coordinates, $T$, where the latter was varied from $0.1$ to $10$, and where the former was evaluated at the maximum time reached in our simulations, $t=100$ (a transient evaluation of $T_\text{eff}$ is presented in the Supplementary Material). All other parameters were taken to be identical to those used in Sec.~\ref{sec:truncations}, and FSSH and MF calculations were performed within their respective reciprocal-space implementation without invoking a basis truncation. Results are shown for both the Holstein and Peierls models. In both cases, $T_\text{eff}$ is seen to converge to $T$ with reasonable accuracy for FSSH, confirming earlier reports of this method satisfying detailed balance \cite{parandekar2005mixed, parandekar2006detailed}. Modest deviations are seen particularly at high temperatures where FSSH yields slight underestimates for $T_\text{eff}$.

Different from FSSH, detailed balance is seen to be dramatically violated within MF dynamics. This method is known to overestimate the effective temperature of a quantum subsystem \cite{parandekar2006detailed, vegte2013calculating}, and indeed, $T_\text{eff}$ is observed to grow beyond $T$. Interestingly, however, we find this growth to continue throughout the entire time interval covered by our simulations, without signs of convergence, and we were therefore only able to provide a lower bound for $T_\text{eff}$ in Fig.~\ref{fig:temp_scan_holstein}. This continued heating of the quantum subsystem is responsible for the excessive spreading of the electronic wavepacket, as it experiences a decreased bias towards the low-energy regions of the Brillouin zone. This in turn explains the transient increase in the cumulative absolute error observed for MF dynamics under Brillouin zone truncations. The violation of detailed balance is significantly more severe than typically found in MF calculations of small quantum subsystems coupled to a quasi-thermodynamic number of classical oscillators \cite{vegte2013calculating}, but can be understood based on the previous demonstration that this violation intensifies with increasing number of closely-spaced quantum states \cite{parandekar2006detailed}.

Whereas detailed balance provides a convenient benchmark for the asymptotic behavior of a method, it does not address the short-time accuracy. In order to assess the latter, we proceed to compare FSSH and MF dynamics with numerically-exact results. Finding reasonably-affordable methods delivering such results for lattice-based models is somewhat challenging. Here, we take advantage of recent efforts of reformulating the hierarchical equations of motion (HEOM) method \cite{tanimura1989time} for the discrete-mode Holstein and Peierls models \cite{liu2014reduced, chen2015dynamics, dunn2019removing}. We specifically employ the implementation \cite{dunn2019removing} of such discrete-mode HEOM in the Parallel Hierarchy Integrator \cite{strumpfer2012open}. Even though HEOM is known to be relatively computationally inexpensive, its unfavorable scaling required us to considerably reduce the number of lattice sites, $N$. Furthermore, within its conventional formulation, discrete-mode HEOM is inherently unstable \cite{dunn2019removing} (which has been addressed in various ways \cite{dunn2019removing, yan2020new}), limiting the time scales reachable by this method. We therefore varied $N$ between 4 and 10 in order to study how the results trend towards larger system sizes, while restricting ourselves to the short-time dynamics before the onset of stabilities. In order to accelerate convergence of the HEOM calculations, we also adjusted the parameters to $J=0.3$, $\omega = 0.3$, $g^{2}=0.25$, and $T=0.5$. For each $N$, the hierarchy depth $L$ was increased until convergence was reached, yielding $L=12$ ($N=4$), 11 ($N=6,8$), and 10 ($N=10$) for the Holstein model and $L=12$ ($N=4,6$) and 10 ($N=8,10$) for the Peierls model. The mixed quantum--classical calculations were again performed in their untruncated reciprocal-space implementations. It should be noted that this comparison is not intended as a comprehensive benchmark of FSSH and MF dynamics for discrete-mode Holstein and Peierls models, which we reserve for a follow-up study, but rather serves to indicate whether either method is capable of yielding reasonable dynamics for the parameters at hand.
\begin{figure}
    \centering
    \includegraphics{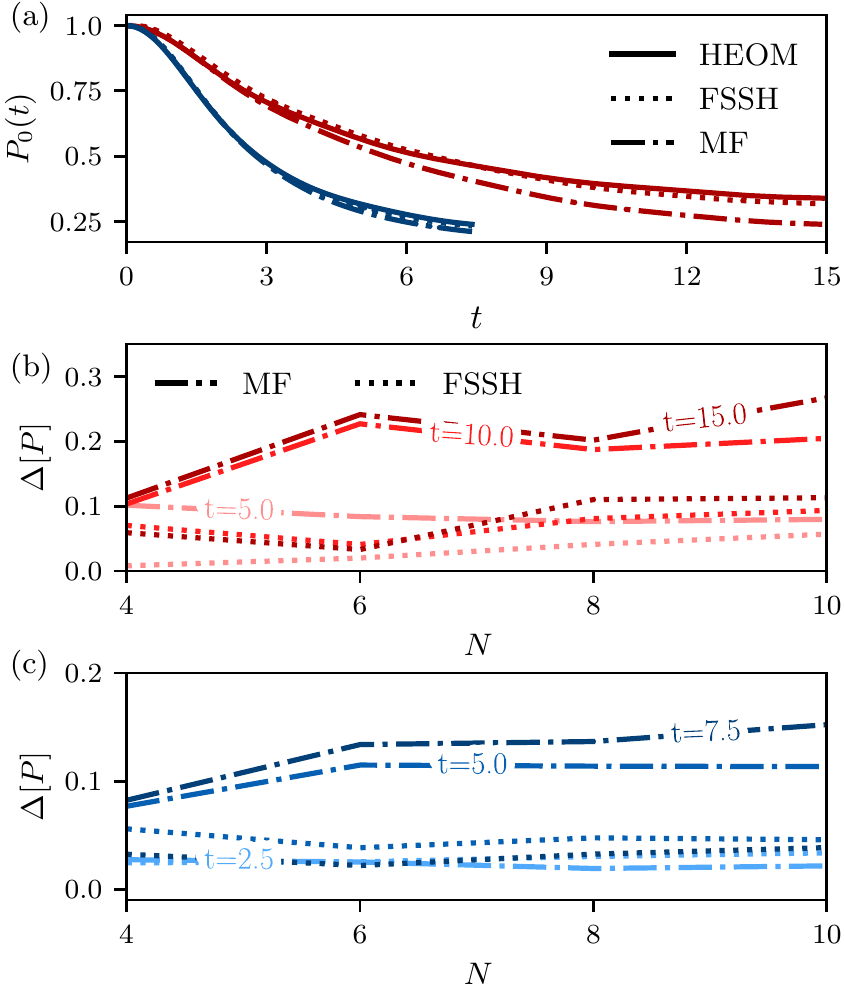}
    \caption{(a) Transient electronic populations $P_{0}(t)$ resulting from MF, FSSH, and HEOM dynamics for $N=10$ site Holstein and Peierls models with $J=0.3$, $\omega = 0.3$, $g^{2}=0.25$, and $T=0.5$. Shown in (a) is $P_{0}(t)$, whereas (b) and (c) present the total absolute error for the Holstein and Peierls models, respectively, shown as a function of $N$ and $t$ (labels added to the curves).}
    \label{fig:benchmark}
\end{figure}

Fig.~\ref{fig:benchmark} presents a comparison between numerically-exact results from HEOM and the outcome from FSSH and MF dynamics for the Holstein and Peierls models. Shown in Fig.~\ref{fig:benchmark} (a) are transient electronic populations at $k=0$ for $N=10$. The results generated by FSSH and MF dynamics are seen to be in fair agreement with the numerically-exact results, even though MF tend to already underestimate the population due to improper thermalization. Fig.~\ref{fig:benchmark} (b) presents the cumulative absolute error at $k=0$ as a function of $N$, shown at different times. For MF dynamics, this error exhibits the expected increase with time, but also with number of lattice sites, $N$. Similar behavior is observed for FSSH, but overall its cumulative absolute error is considerably smaller than that of MF dynamics. Interestingly, for the Peierls model, the error is seen to decrease in going from $t=5.0$ to $t=7.5$ as FSSH begins to equilibrate to the appropriate thermal distribution, reflective of its correct asymptotic behavior.
\section{Conclusions and outlook}\label{sec:conclusions}

In summary, we have presented a formulation of FSSH where the quantum and classical equations of motion are solved entirely within a reciprocal-space representation. Using a tight-binding model involving Holstein-type and Peierls-type electron--phonon couplings, this approach is shown to be formally equivalent to the conventional real-space formulation of FSSH. Where the real-space and reciprocal-space representations differ is in their potential for basis truncations. Molecular systems typically are best truncated in real space, due to the local nature of electronic excitations, whereas reciprocal-space truncations are preferable for band-like materials \cite{qiu2016screening, tempelaar2019many}, where electronic excitations scatter between Bloch-like states as a result of interactions with phonons. As such, the two representations span the two opposites in optimally describing hopping versus band-like transport.

Compared to MF (or Ehrenfest) dynamics, for which a reciprocal-space formulation was proposed in a previous study by the authors \cite{krotz2021reciprocal}, the ability to perform basis truncations in reciprocal space is affected by the improved thermalization properties of FSSH. For a single electronic carrier under Holstein coupling, the resulting wavepacket is thermally biased to the band minimum located at zero momentum, allowing for effective truncations of the Brillouin zone around $k=0$. Sizable Peierls coupling, however, gives rise to band minima at $k=\pm\pi/2$ \cite{zhao1994on, chen2011on}, as a result of which the electronic wavepacket splits in two branches. Although this complicates a Brillouin zone truncation around $k=0$, it is conceivable that more elaborate truncation schemes may still be effective, which would be an interesting topic of future inquiry.

Whereas our FSSH results are seen to obey detailed balance to within reasonable accuracy for both Holstein and Peierls models, we find MF dynamics to dramatically overestimate the effective temperature within the quantum subsystem for both models, which is caused by the large number of closely-spaced adiabatic states arising in band-like problems. When compared to numerically-exact results, both FSSH and MF dynamics are seen to reach reasonable accuracy at early times, before thermalization sets in. For FSSH we thus find the performance to be promising both in the short-time and asymptotic limits. It should, however, be noted that the application of FSSH to discrete-mode Holstein and Peierls lattices has remained underexplored compared to its utility in small electronic systems coupled to a quasi-thermodynamic number of classical oscillators, with a few recent exceptions noted \cite{wang2013flexible, wang2014simple, wang2015mixed, qiu2018crossing, sun2019impurity, sun2020polaron, prodhan2020design, sun2021surface, huang2021unified}, and there is much left to be done in order to understand its potentials and pitfalls.

The reciprocal-space electronic populations obtained within our FSSH formulation are derived from an adiabatic density matrix that inconsistently treats populations (based on active surfaces) and coherences (based on wavefunction coefficient) -- see Eq.~\ref{eq:DM_eigenbasis}. Although such treatment has previously been shown to work well in many cases \cite{tempelaar2013surface, landry2013correct, chen2016on, tempelaar2018generalization}, it would be of interest to consider reciprocal-space formulations based on recently-proposed generalizations of FSSH that consistently express the entire adiabatic density matrix in terms of active surfaces \cite{wang2015fewest, tempelaar2018generalization}. Not only could this improve the accuracy in certain regions of parameter space \cite{tempelaar2018generalization}, but it also allows for a straightforward extension to simulate the nonlinear spectroscopy of crystalline materials \cite{tempelaar2019many}. It would also be of interest to combine reciprocal-space FSSH with a ring-polymer approach to account for quantum effects of the involved phonon modes \cite{shushkov2012ring, shakib2017ring, lu2017path}.

Lastly, it is worth mentioning that the equations of motion found in reciprocal-space FSSH share the same structure as their real-space equivalents. It would therefore be straightforward to include the formalism presented here in existing (real-space) implementations of surface hopping. The ease of implementation, combined with the opportunities for effective basis truncations and a direct compatibility with band structure calculations, renders reciprocal-space surface hopping an attractive method for the simulation of nonequilibrium phenomena in crystalline materials.
\section*{Supplementary material}

See Supplementary Material for mean-field dynamics results and transient fittings of the effective temperature of the quantum subsystem.

\section*{Data availability}

The data that support the findings of this study are available from the corresponding author upon reasonable request.

\section{Acknowledgements}

Research reported in this publication was supported, in part,
by the International Institute for Nanotechnology at Northwestern
University.

\bibliography{Bibliography}

\end{document}


\title{Supplementary Material for ``A Reciprocal-Space Formulation of Surface Hopping''}

\author{Alex Krotz}
\author{Roel Tempelaar}
\email{roel.tempelaar@northwestern.edu}

\affiliation{Department of Chemistry, Northwestern University, 2145 Sheridan Road, Evanston, Illinois 60208, USA}

\maketitle

\begin{figure}[h!]
    \centering
    \includegraphics[scale=1.0]{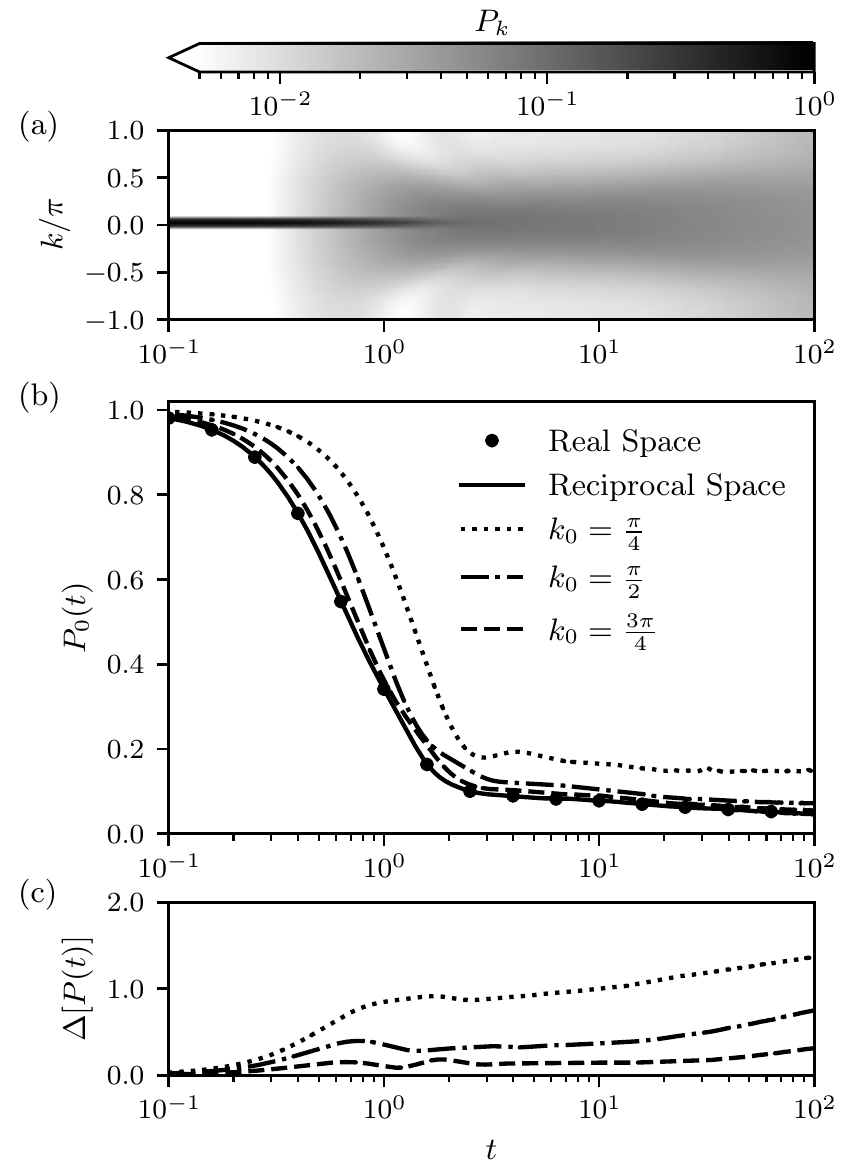}
    \caption{Same as Fig.~2 of the main text, but for MF dynamics. Data reproduced from Ref.~\citenum{krotz2021reciprocal}.}
    \label{fig:MFD_holstein}
\end{figure}

\begin{figure}[h!]
    \centering
    \includegraphics[scale=1.0]{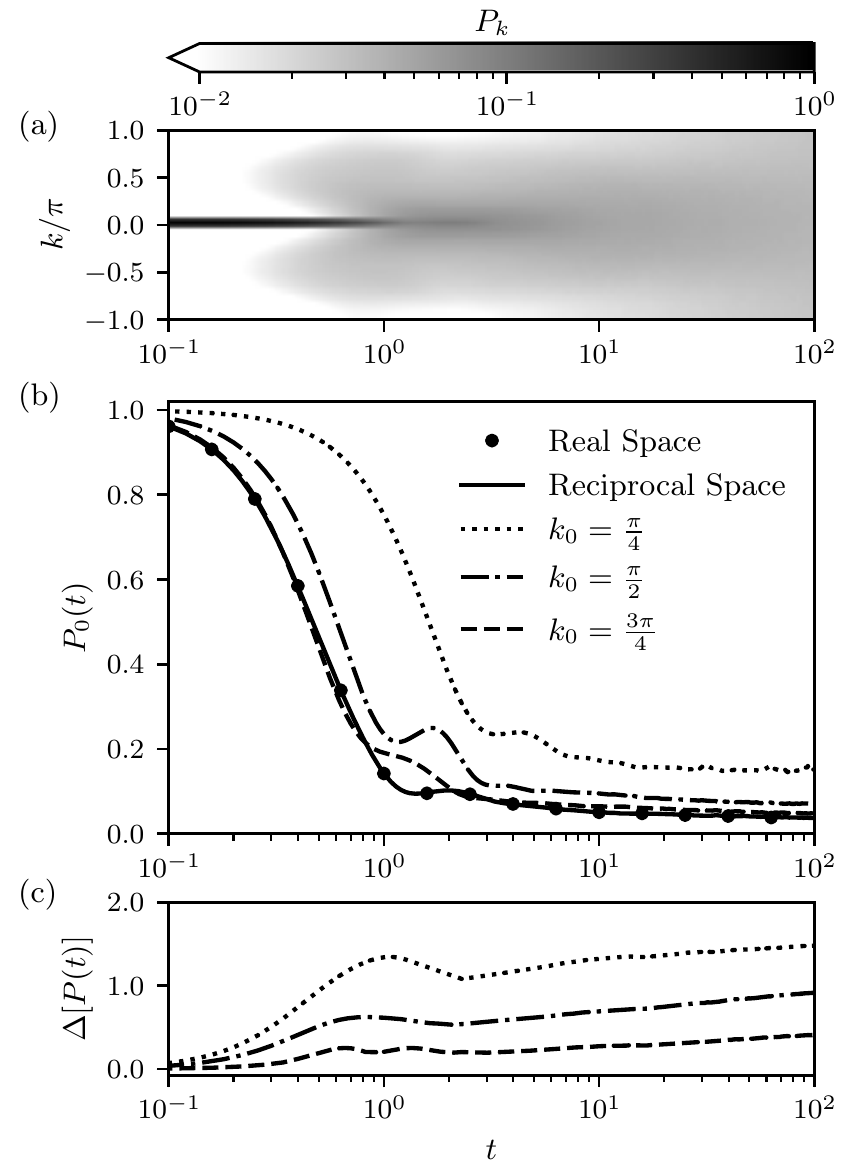}
    \caption{Same as Fig.~3 of the main text, but for MF dynamics. Data reproduced from Ref.~\citenum{krotz2021reciprocal}.}
    \label{fig:MFD_peierls}
\end{figure}

\begin{figure}[h!]
    \includegraphics[scale=1.0]{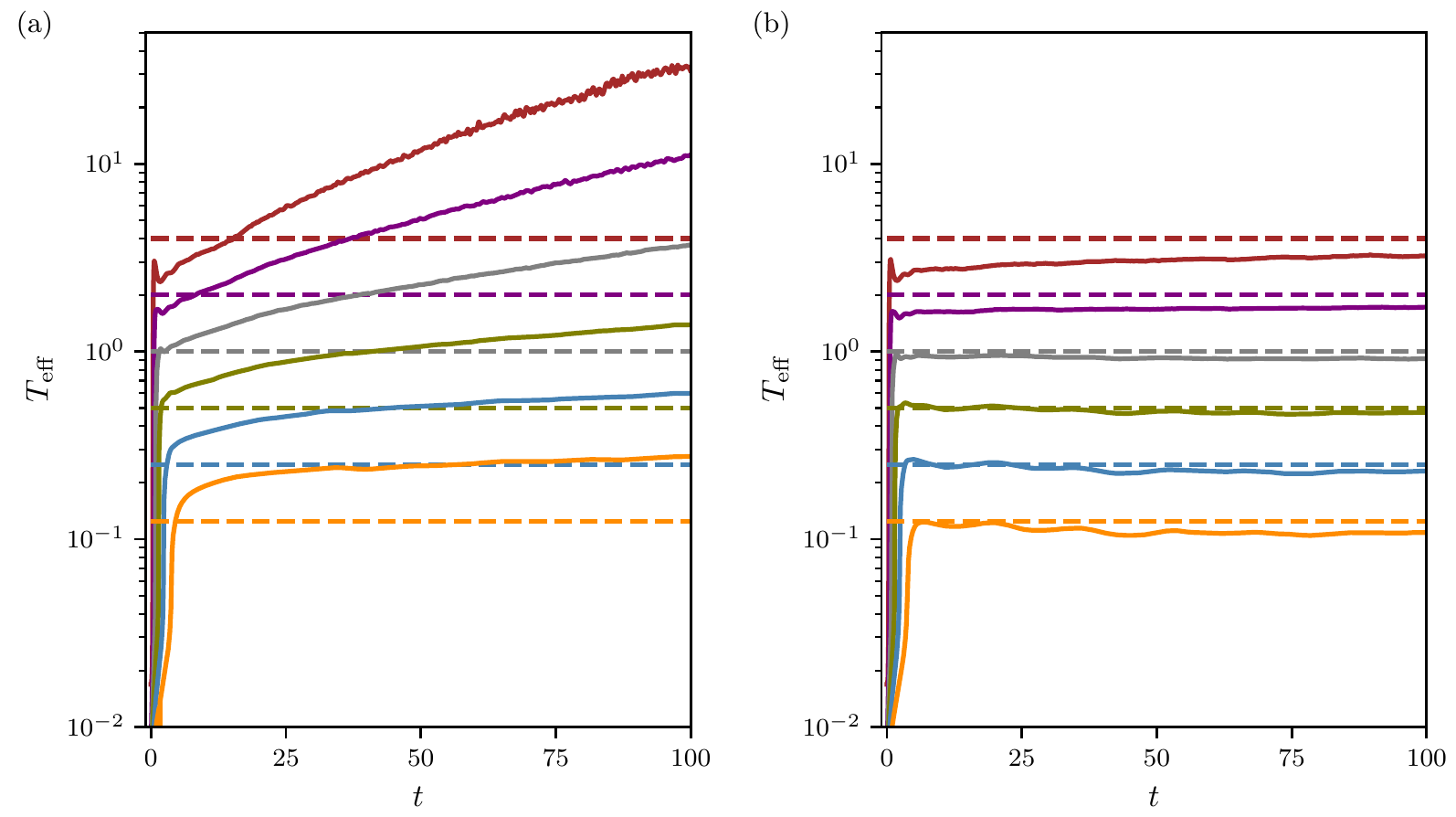}
    \caption{Time-dependent effective temperature of the quantum subsystem, $T_\text{eff}$, obtained through the fitting procedure described in the main text. Shown are results for the $N=\varN$ site periodic Holstein model with $J=1.0$, $\omega = 0.2$, and $g^{2}=5.0$, and with varying temperature of the classical coordinates, $T$ (shown as horizontal dashes). Results obtained through MF and FSSH dynamics are shown in (a) and (b), respectively.}
    \label{fig:thermalize_ed_fssh}
\end{figure}

\begin{figure}[h!]
    \centering
    \includegraphics{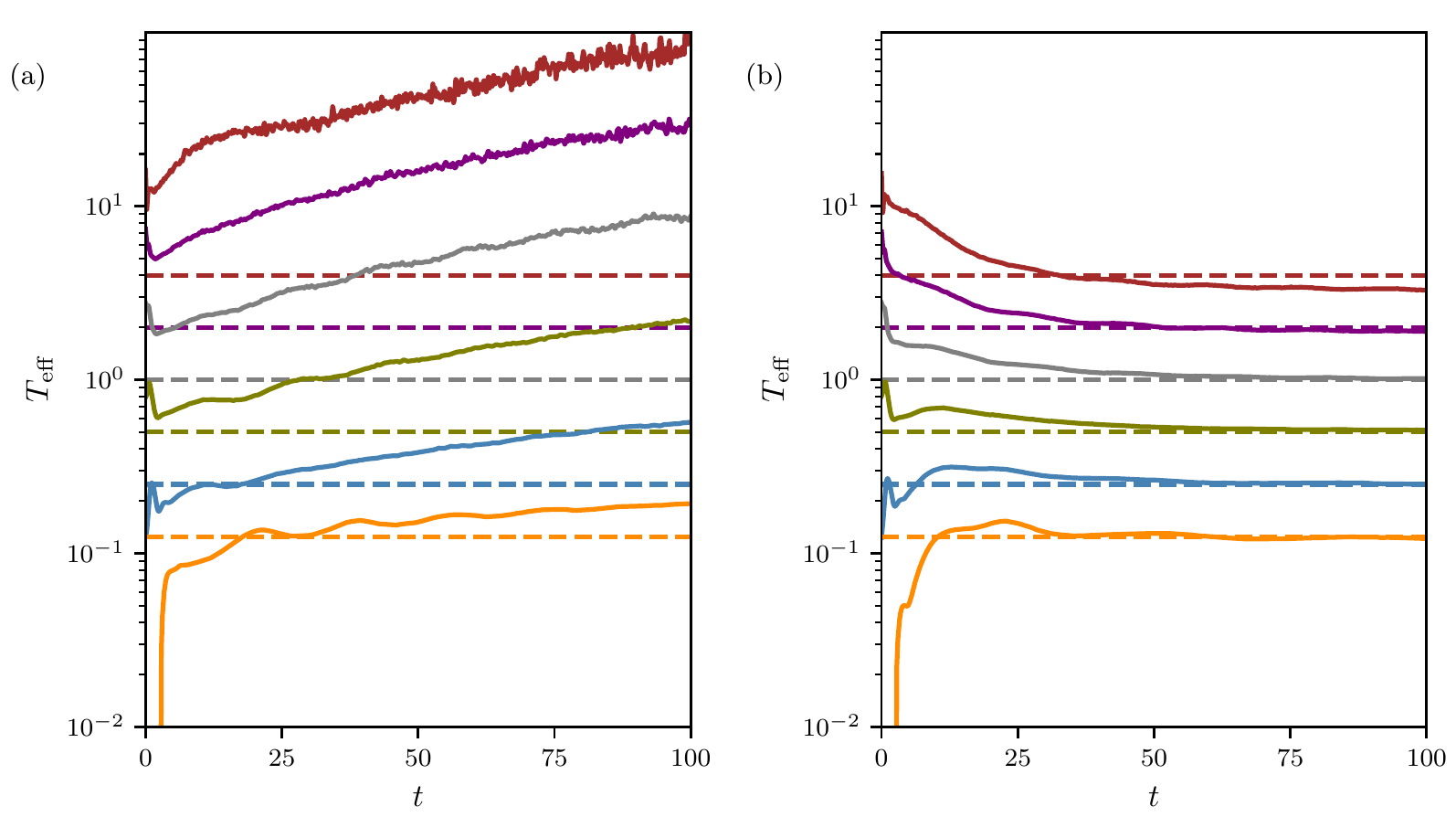}
    \caption{same as Fig.~\ref{fig:thermalize_ed_fssh} but for the $N=\varN$ site periodic Peierls model.}
\end{figure}

\clearpage

\bibliography{Bibliography}